\documentclass{webofc}

\usepackage[varg]{txfonts}   
\usepackage{hyperref}
\usepackage{url}
\hypersetup{colorlinks=true,citecolor=blue,urlcolor=blue,linkcolor=blue}
\usepackage{subcaption}
\usepackage{graphbox}

\begin{document}
\pagestyle{fancy}
\fancyhead[R]{\thepage}

%
\title{Overview of recent results from the STAR experiment}
%
%

\author{\firstname{Isaac} \lastname{Mooney}\inst{1,2}\fnsep\thanks{\email{isaac.mooney@yale.edu}} for the STAR Collaboration
}

\institute{Wright Laboratory, Yale University, New Haven, CT 
\and
           Brookhaven National Laboratory, Upton, NY
          }

\abstract{We highlight the STAR experiment's recent measurements on electromagnetic and hard probes of nuclear collisions, which inform the field's understanding of these physical phenomena and by extension QCD. Results on vector meson production from the high electromagnetic fields in glancing heavy-ion collisions are presented. Observables related to jets, high-momentum hadrons, heavy quarks and quarkonia in vacuum and their modification in head-on heavy-ion collisions are also presented. Studies using electromagnetic probes of the medium created in these collisions are presented as well. Finally, we conclude and give an outlook for STAR data-taking and measurements in the coming years.\par
}
\maketitle

\section{Introduction}
\label{intro}

	
At the earliest times in a heavy-ion collision, hard scattering between partons in each nucleus may occur, which results in outgoing high-transverse-momentum products. These products, either highly virtual light partons or heavy quarks, then start to evolve via a shower of collimated radiation (resulting in ``jets''). 
As the quark-gluon plasma (QGP) thermalizes, these color charges begin to interact, mostly via gluon bremsstrahlung, causing energy deposition into the medium and resulting in a modification of yields and angular broadening. The response of the medium to these interactions may also be measured as a modification of jet yield and substructure. Gluon Compton scattering and $q-\bar{q}$ annihilations in the evolving medium will also occur, and the resulting photon or dilepton pair will escape the medium relatively unscathed due to their lack of color charge, making access to the temperature of the medium possible for experimenters. Conversely, $q-\bar{q}$ bound states may be generated or dissociate in the QGP. Finally, the medium and the strongly-interacting probes will hadronize and undergo decays before the final-state particles are measured at asymptotically late times. In glancing, or (ultra-)peripheral, collisions, the QGP is either not formed or has minimal spatiotemporal extent, resulting in correspondingly minimal in-medium contributions mentioned above. This allows for the study of extreme electric and magnetic fields created by near-lightspeed passage of electric charges from the colliding nuclei, and contributions to the final measurements other than the QGP (such as cold nuclear matter effects). STAR has recently made measurements which highlight all of these phases of heavy-ion collisions, discussed below. \par
In these proceedings, we highlight STAR's recent measurements presented at the 12th International Conference on Hard and Electromagnetic Probes of High-Energy Nuclear Collisions in Nagasaki, Japan. Flavor correlators of leading hadrons, and charged energy correlators are discussed in the context of an improved understanding of the non-perturbative hadronization process within jets. Jet $v_{1}$ is presented as a novel approach to revealing the path-length dependence of energy loss. 
Ratios of proton and pion yields are measured in order to search for a possible medium-response to this jet energy deposition, in the form of enhanced coalescence of baryons relative to mesons. The radius dependence of charm-jet nuclear modification is studied to determine whether this medium-response contribution and recovery of lost energy, or increased energy loss, would be the dominant contribution for heavy quarks and their associated radiation in the QGP. $J/\psi$ production in vacuum and its modification in the medium is studied with self-normalized yields and nuclear modification factors respectively, where the former focuses on the multiplicity dependence of production, and the latter focuses on the energy dependence due to primordial and regeneration contributions. Additionally, the first measurement of sequential suppression of charmonium at RHIC is presented. We also report reliable QGP temperature extractions with measurements of thermal dielectrons. Lastly, photon-induced vector boson decay asymmetries are presented as a potential reliable method for reaction plane estimation. Finally, we conclude and present an outlook toward the future of STAR data-taking and data analysis.

\section{Detector}
\label{sec:det}

RHIC is a highly versatile collider, capable of offering beams of many ion species, from hydrogen to uranium, and ranging from $\sqrt{s_{\mathrm{NN}}} = 3 \mathrm{\ GeV}$ in fixed-target mode to $200 \mathrm{\ GeV}$ ($510 \mathrm{\ GeV}$ for proton-proton (pp) collisions). Partly because of this, STAR is also a highly versatile detector, with numerous subsystems capable of measurements across a broad kinematic range. The Time Projection Chamber (TPC) at $|\eta| < 1$ provides tracking and momentum determination of charged particles, as well as a centrality measure via selections on the multiplicity of these charged tracks. The Barrel Electromagnetic Calorimeter (BEMC) at $|\eta| < 1$ measures energy deposits of photons, electrons, and charged hadrons, and additionally acts as a fast online trigger. The Time of Flight (TOF) detector at $|\eta| < 0.9$ identifies particles by their speed, and is also useful for pileup mitigation. Lastly, a silicon tracker close to the beamline called the Heavy Flavor Tracker (HFT), located at $|\eta| < 1$, was installed from 2014 to 2016 and allowed for precise access to displaced vertices produced by decays of heavy particles. At larger rapidities are the Vertex Position Detector (VPD) at $4.24 < |\eta| < 5.1$ and Zero Degree Calorimeter (ZDC) at $18\mathrm{\ m}$, both of which can be used as minimum bias triggers, while the VPD is also used for vertex reconstruction with precise timing and therefore position resolution, and the ZDC is also used for luminosity monitoring. \par

\section{High momentum hadrons and correlations}
\label{sec:hadrons}

\begin{figure*}
\centering
\begin{subfigure}[b]{0.5\textwidth}
        \centering
        \includegraphics[align=c,width=5cm,clip]{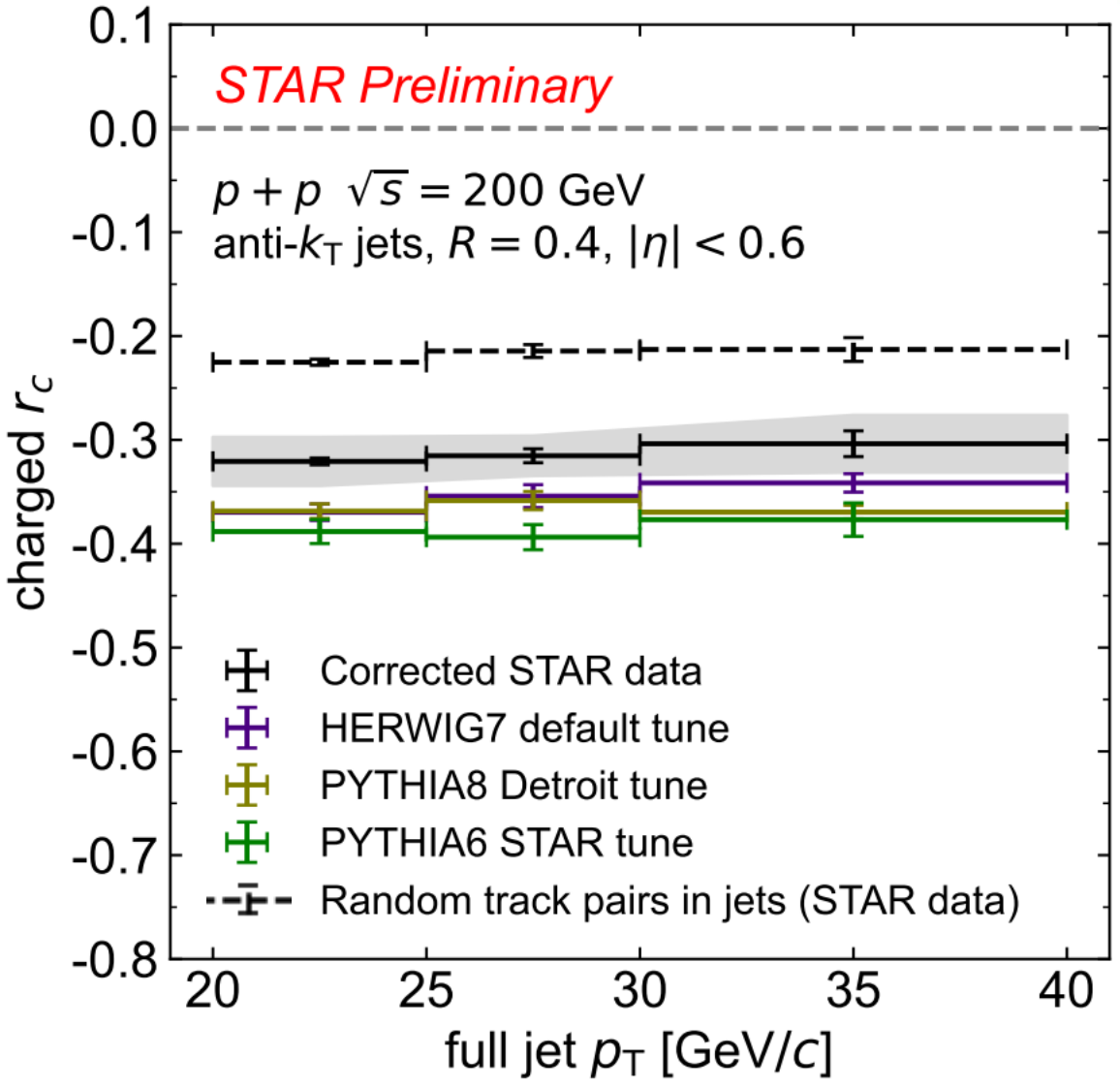} 
    \end{subfigure}%
    \begin{subfigure}[b]{0.5\textwidth}
        \centering
        \includegraphics[align=c,width=6cm,clip]{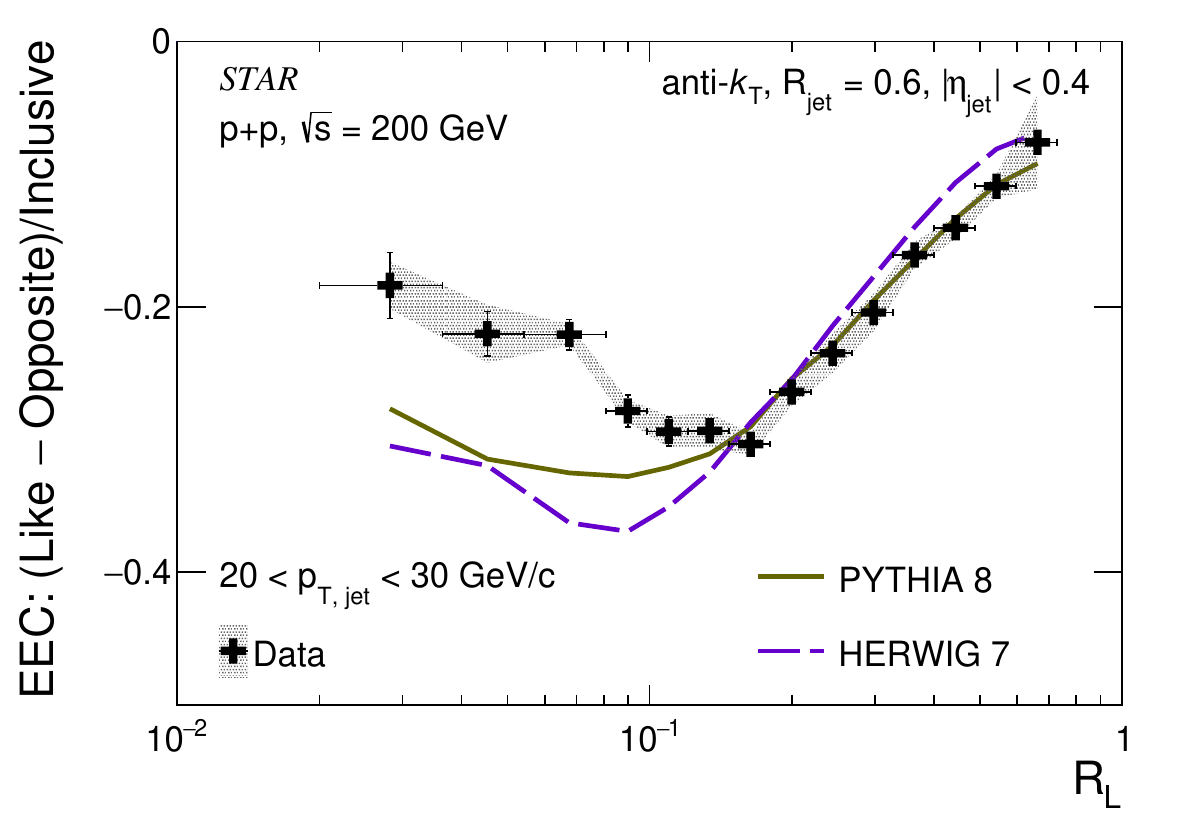} 
    \end{subfigure}%
\caption{Left: Corrected measurement, as a function of jet transverse momentum ($p_{\mathrm{T}}^{\mathrm{jet}}$) of the charge correlator ratio in pp collisions using leading track pairs (black solid lines) compared to the same observable defined on random tracks (black dashed lines), and Monte Carlo models (colored solid lines). Right: Corrected measurement, as a function of angular separation, of the charge-selected energy correlator ratio in pp collisions (black markers) compared to Monte Carlo models (colored lines).}
\label{fig:youqiandrew}
\end{figure*}

The flavor correlator, $r_{c}$, of the leading two charged hadrons in a jet has recently been proposed~\cite{YT} within the Electron-Ion Collider (EIC) kinematic environment as an observable which can highlight the hadronization mechanism within jets. It is defined as
\begin{equation}
r_{c}(X) = \dfrac{\mathrm{d}\sigma_{h_{1}h_{2}}/\mathrm{d}X - \mathrm{d}\sigma_{h_{1}\bar{h}_{2}}/\mathrm{d}X}{\mathrm{d}\sigma_{h_{1}h_{2}}/\mathrm{d}X + \mathrm{d}\sigma_{h_{1}\bar{h}_{2}}/\mathrm{d}X}.
\end{equation}
From this definition, one sees that the range of the observable is -1 to 1. The lower bound here corresponds to a perfect opposite charge correlation for the leading two constituents. If hadronization occurred exclusively through string fragmentation, this would be the expected result. The $r_{c}$ of random tracks from STAR jets 
is roughly -0.2. The data would then be expected to fall somewhere between -1 and -0.2\footnote{This is true even if the hadronization mechanism is perfectly string-like, since resonance decays may dilute the correlation.}. The measurement was recently performed in pp collisions for the first time by STAR \cite{YouqiTP}, as a function of transverse momentum ($p_{\mathrm{T}}$) of full (charged+neutral) jets (Fig.~\ref{fig:youqiandrew} (left)). We observe an $r_{c}$ which is independent 
of $p_{\mathrm{T}}^{\mathrm{jet}}$, at roughly -0.3. Because \textsc{Herwig} and \textsc{Pythia} have different hadronization mechanisms, the data are compared to Monte Carlo models \textsc{Herwig}7, \textsc{Pythia}6, and \textsc{Pythia}8. At the moment, extracting a physics conclusion from these comparisons is difficult, as a full \textsc{Herwig}7 tune to the RHIC environment is not yet published \cite{HerwigTune}, and resonance decays have a different contribution in each model.
However, all models predict more charge correlation than shown in the data. An extension of this measurement to heavy-ion collisions to search for potential modification to the hadronization process in jets is ongoing \cite{YouqiTP}.\par 
Correlators of energy flow operators, called ``energy correlators'', in the small-angle limit have recently been measured by STAR in pp collisions \cite{andrew}. The two-point correlator is defined on all pairs of (charged) constituents in a jet as 
\begin{equation}
\mathrm{Normalized\ EEC} = {\dfrac{1}{\mathfrak{C}} \dfrac{\mathrm{d}\mathfrak{C}}{\mathrm{d}R_{L}} }, \mathrm{\ where\ }\mathfrak{C} = \sum_{\mathrm{jets}}\sum_{i\neq j}\dfrac{E_{i}E_{j}}{p_{\mathrm{T,jet}}^{2}} 
\end{equation}
and extended to three-point correlations by examining all triplets, with the energy weight extended in the obvious way, and typically projected onto the longest intra-particle distance of the triplet, $R_{L}$. STAR presents preliminary measurements extending the results to this three-point correlator, as well as measurements of the charge-dependent correlator, which is defined by including a factor of each constituent's charge in the energy weight. The main purpose of the latter observable is to examine the charge flow of hadronization,  similarly to the $r_{c}$ measurement mentioned above. The data are shown in Fig.~\ref{fig:youqiandrew} (right) as a ratio to the inclusive two-point energy correlator. The data prefer more opposite-sign than like-sign pairs across the entire $R_{L}$ domain. This effect increases from the largest angles (roughly corresponding to the perturbative and partonic early stage) to the partonic-to-hadronic transition regime, then decreasing toward the smallest angles in the hadronic regime. In the large-angle regime, the Monte Carlo models \textsc{Pythia}8 and \textsc{Herwig}7 are consistent with data. However, in the small-angle hadronic regime, they significantly underpredict the data. This behavior is consistent with that seen in the $r_{c}$ measurement mentioned above. These measurements are also being extended to heavy-ion collisions, where it has been predicted \cite{AndrewTP} that there may be a modification to the scaling behavior seen in pp collisions at large angles (early times)\footnote{Measurements in pp collisions are appropriate baselines, as cold nuclear matter effects on jets are negligible~\cite{DaveVeronica}.}. \par

\section{Jet modification and medium response}

\begin{figure*}
\centering
\includegraphics[width=7cm,clip]{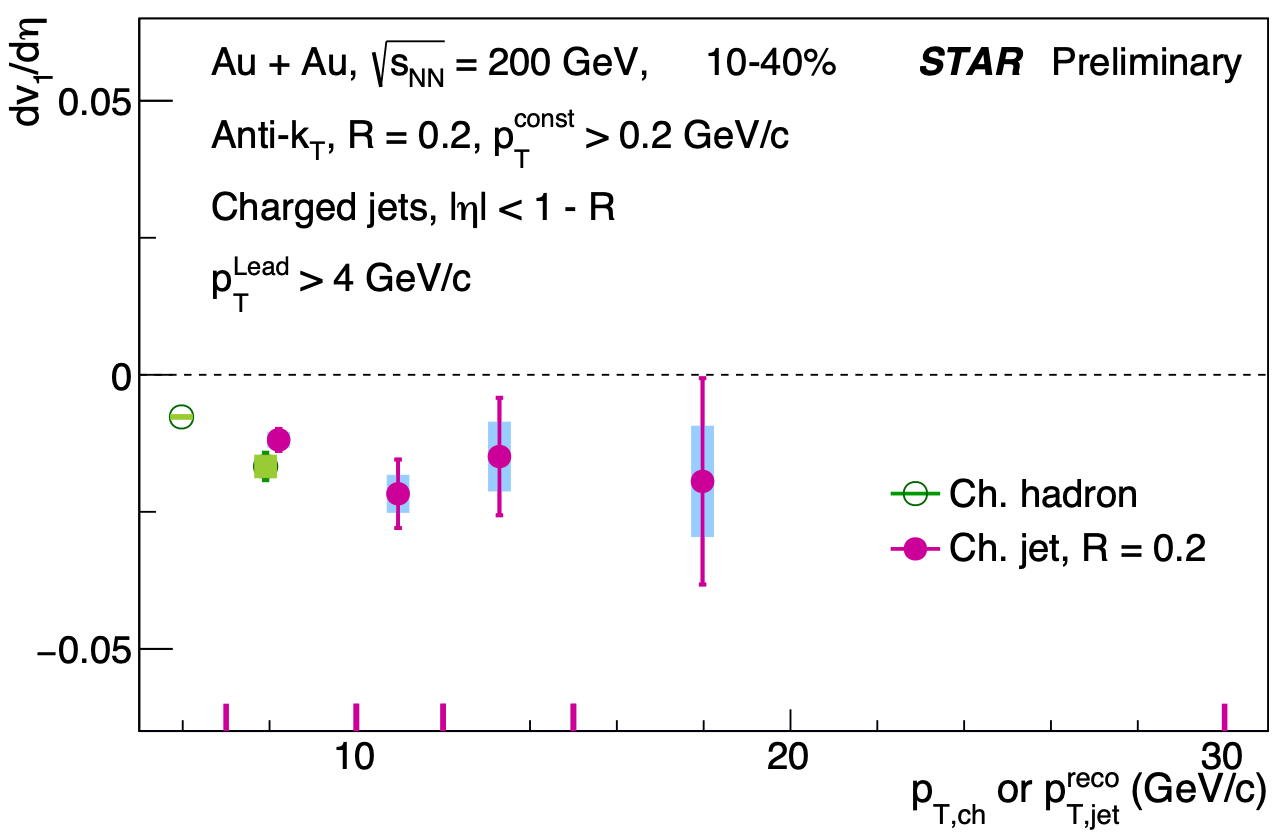}
\caption{Measurement of the slope of jet (magenta solid circles) and charged-hadron (green open circles) $v_{1}$ versus pseudorapidity, as a function of $p_{\mathrm{T}}$ in Au+Au collisions.}
\label{fig:sooraj}
\end{figure*}

As for measurements at STAR in heavy-ion collisions, a novel observable to probe the path-length dependence of energy loss is the jet $v_{1}$. Because the bulk is tilted in these collisions, while the hard process production profile is isotropic, at large positive (negative) rapidities we may expect to observe the result of a longer path through the QGP for jets traveling in the positive (negative) $x$-direction. This would result in a finite $v_{1} = \langle \cos(\phi - \Psi_{\mathrm{RP}}) \rangle$, where $\Psi_{\mathrm{RP}}$ is the reaction-plane angle, and $\phi$ is the azimuthal angle of the jet. STAR observes a significant jet $v_{1}$ (see Fig.~\ref{fig:sooraj}) for all jet radii and $p_{\mathrm{T}}$ studied, in Au+Au and isobaric (Zr+Zr and Ru+Ru) collisions. Work is ongoing to connect this to path-length dependence, as well as to enhance the signal using event-shape engineering with multiplicity fluctuations~\cite{SoorajTP}. \par
As jets traverse the medium, they not only lose momentum, but also induce a response from the medium which may enter the final-state jet. This response may include coalescence of medium partons which would be more likely baryonic than the hadrons resulting from vacuum fragmentation. This was predicted for LHC kinematics~\cite{AMPT_LHC} to cause an enhancement of the $p/\pi$ ratio in Pb+Pb collisions that is most pronounced at large distances from the jet axis in most central collisions, and has recently been measured at STAR~\cite{GabeTP}. We observe no modification to this ratio in jets in central Au+Au collisions compared to pp collisions, despite a large modification of the same ratio in the analogous comparison for inclusive particles \cite{inclusive_baryonmeson}, suggesting that fragmentation is still the dominant mechanism for hadron production in jets in heavy-ion collisions at STAR. The analysis was recently extended from a constituent $p_{\mathrm{T}}$ threshold of $3$ to $2\mathrm{\ GeV}$ to attempt enhancing the medium-response signal, which results in larger uncertainties from the correspondingly increased background contribution, especially for the largest radius studied ($R = 0.4$). Work is ongoing to reduce systematic uncertainties. \par
\section{Heavy quarks and quarkonia}
Heavy quarks are excellent probes of the QGP because of their large mass, which causes production to occur early in the evolution of the collision; aids calculability in perturbative QCD; and limits the likelihood of thermalization. By examining heavy-quark-jets (containing at least one hadron with a heavy quark), one also retains the medium-induced radiation associated with the passage of the original heavy-quark. STAR recently measured the nuclear modification factor of $D^{0}$-meson ($c\bar{u}$) jets, comparing central and peripheral Au+Au collision yields at $\sqrt{s_{\mathrm{NN}}} = 200 \mathrm{\ GeV}$, where it was observed for example that jets containing hard-fragmenting (high-$z$) $D^{0}$s were suppressed in central collisions. Recently, the measurement was extended to consider the jet radius dependence of this nuclear modification. The naïve expectation is that wider jets have more medium interaction sites causing more energy loss, which would cause larger suppression of $R = 0.4$ jets compared to $R = 0.2$ jets, while they would recover more of the lost energy in a larger cone and have more potential to recover medium response, leading to opposite behavior. The STAR preliminary measurement finds no $R$-dependence within large uncertainty, suggesting within current precision that the contribution of these effects is similar in magnitude. The results are consistent with models~\cite{LIDO,CCNU} which also predict minimal $R$-dependence. This measurement will be extended to include the $D^{0}$-jet generalized angularities, which are IRC-safe jet substructure observables.\par 
In the medium, $c\bar{c}$ pairs can dissociate due to Debye screening from the thermal bath. However, they can also be regenerated in the medium. At low energies, the former effect is thought to be much more prevalent, meaning that RHIC offers the opportunity to study a physical regime distinct from the LHC. STAR has recently taken a wealth of new Au+Au data from the Beam Energy Scan II (BES-II) program across a large range of $\sqrt{s_{\mathrm{NN}}}$, with roughly ten times the statistics of the BES-I program. The recently measured $J/\psi$ nuclear modification factor in central Au+Au collisions at center-of-mass energies ranging from $14.6$ to $27\mathrm{\ GeV}$ shows that there is a suppression of the $J/\psi$ compared to pp collisions, consistent with a transport model~\cite{Jpsi_BES_calc} including primordial and regeneration production, with minimal energy dependence up to at least top RHIC energy. \par 
Additionally, the excited states of the $J/\psi$ may dissociate more easily in the medium, given their lower binding energies which make them more susceptible to Debye screening. In this way, they are sensitive to the temperature of the QGP, as a hotter medium should cause dissociation of more tightly bound states. STAR recently made the first observation of sequential suppression of charmonium at RHIC, with a double ratio of the $\psi(2S)$ to $J/\psi$ cross section compared between $200 \mathrm{\ GeV}$ isobar and pp collisions being significantly below unity. Both charmonium analyses are being finalized for publication~\cite{WeiTP}.\par
Although pp collisions are used as a baseline for observation of modification of charmonium production by the QGP, charmonium production in vacuum is still not fully understood. It is thought that secondary partonic interactions (MPI) may play a role, especially at higher multiplicities \cite{Jpsi_MPI}, as well as percolation of color strings \cite{Jpsi_percolation}. These expectations have been tested in the past by measuring the self-normalized yield of $J/\psi$ as a function of the self-normalized multiplicity, where a faster than linear rise has been observed \cite{ALICE_Jpsi_7000, STAR_Jpsi_200, ALICE_Jpsi_13000}. Recently STAR has made a preliminary measurement on data from $510 \mathrm{\ GeV}$ collisions~\cite{BrennanTP}. These data are consistent with the previous $200 \mathrm{\ GeV}$ results, with a finer multiplicity selection, and an extension to higher self-normalized multiplicity (which is measured in this case by the TOF, and not corrected for detector effects, which are small). The uncertainties in the highest multiplicity intervals are large, but there may be a hint of a different functional trend between the RHIC and LHC data, which will be an interesting puzzle to resolve. Before publication, this analysis will include a multiplicity correction, to ensure that the comparison is valid. \par

\section{Electromagnetic and electroweak probes}
Color-charged probes of the medium are extremely useful tools for elucidating its microscopic properties because they interact as they pass through it. Minimally-interacting electroweak probes have also been used to extract properties of the medium that are unaffected by these final-state interactions due to their mean free path being longer than the size of the medium. STAR has presented results recently using thermal dielectrons, which additionally are unaffected by the blueshift that complicates direct-photon-enabled QGP temperature extraction~\cite{ChenliangTP}. Electron pairs with moderate invariant mass (falling in the IMR, or intermediate-mass region of an invariant mass spectrum), are thought to be dominantly produced via quark-anti-quark annihilation in the early partonic phase, while those with low mass (in the LMR) are thought to be produced by $\rho$ resonances in the later stage of medium evolution close to the pseudo-critical temperature, $T_{\mathrm{PC}}$. By removing background sources such as Dalitz decays from the data by extracting the result of a cocktail fit, the remaining excess from each invariant-mass region can be fit with a different functional form to extract the temperature. In a preliminary result using isobar collision data, the extracted temperature from the IMR is $293\pm 11(\mathrm{stat.})\pm 27 (\mathrm{syst.}) \mathrm{\ MeV}$, which is well above $T_{\mathrm{PC}}$, as expected. $T_{\mathrm{LMR}}$ on the other hand is $199\pm 6(\mathrm{stat.})\pm 13 (\mathrm{syst.}) \mathrm{\ MeV}$, in slight tension with the expectation from a phase-transition-dominated production. This analysis is being finalized for publication.\par
STAR also recently made a preliminary measurement of thermal dielectron yields using BES-II data, with temperature extractions from $\sqrt{s_{\mathrm{NN}}} = 14.6$ and $19.6\mathrm{\ GeV}$ using a similar approach, giving $T_{\mathrm{LMR}} = 183\pm 25(\mathrm{stat.})\pm 21 (\mathrm{syst.}) \mathrm{\ MeV}$ and $T_{\mathrm{LMR}} = 168\pm 13(\mathrm{stat.})\pm 15 (\mathrm{syst.}) \mathrm{\ MeV}$, respectively. These results are consistent with $T_{\mathrm{PC}}$, suggesting that emission occurs predominantly near the phase transition. This analysis will be updated by reducing the photonic conversion background to improve statistics, and more energies will be analyzed.\par

\section{Nuclear PDFs, saturation, and early-time dynamics}

\begin{figure*}
\centering
\includegraphics[width=6cm,clip]{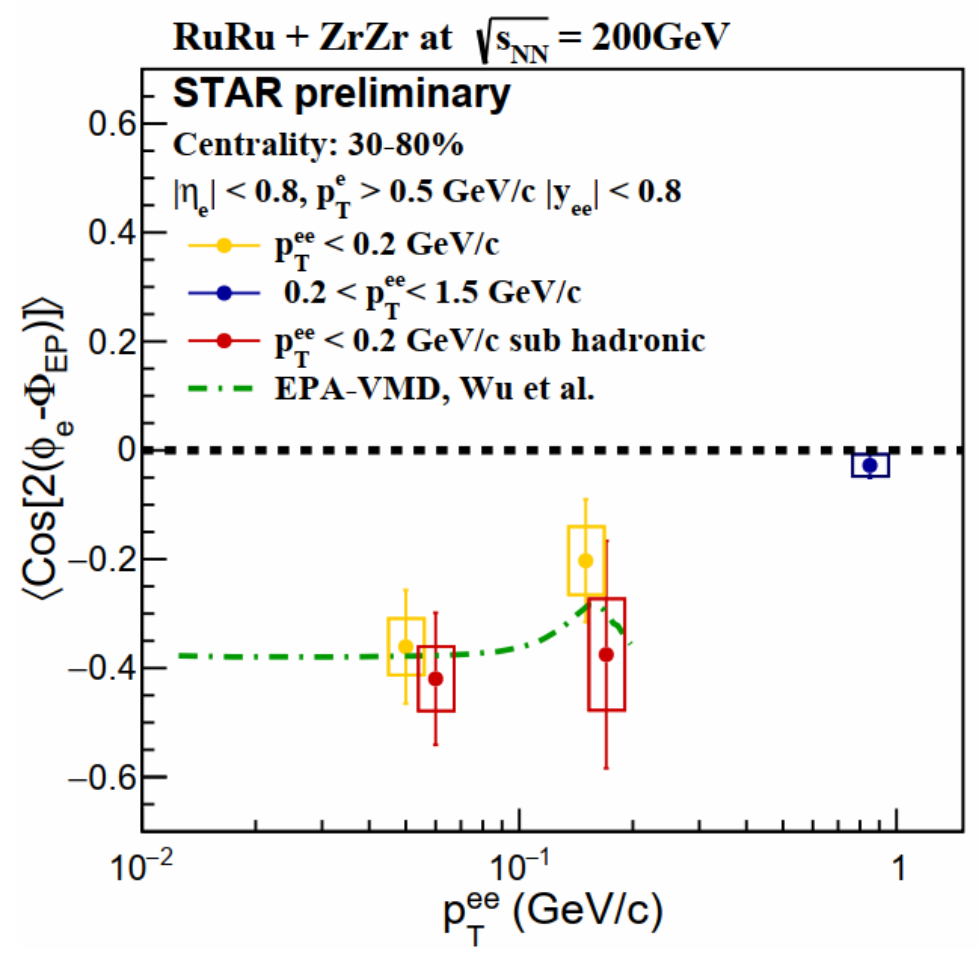}
\caption{Measurement in isobar collisions of the azimuthal modulation of $J/\psi$ decay products with respect to the event plane, as a function of dielectron transverse momentum, compared to an equivalent photon approximation - vector meson dominance calculation (green dot-dashed line). Orange (red) markers are before (after) subtraction of the hadronic contribution estimated from an extrapolated Tsallis function fit.}
\label{fig:kaiyang}
\end{figure*}

While the QGP is an excellent testbed for hot QCD, heavy-ion collisions which produce minimal if any QGP are also useful tools. They refine our understanding of the baseline for hot QCD effects, and create a cleaner environment which still exhibits extreme conditions -- for example, of the electromagnetic field. As nuclei travel past each other at almost the speed of light, they exert strong electric and magnetic fields almost perpendicular to their direction of travel, which can be approximated as a flux (along the direction of travel) of linearly polarized ``Weizsäcker-Williams'' photons in the equivalent photon approximation (EPA) \cite{EPA-VMD}. One of the photons from the projectile nucleus can interact with the target nucleus to produce a vector meson, such as a $J/\psi$. 
The produced $J/\psi$ inherits the linear polarization from the photon, which may then persist to the $e^{+}e^{-}$ pair created when it decays via the dielectron channel, causing an azimuthal anisotropy of the products, with a direction related to the collision's event plane. This was observed recently~\cite{KaiyangTP} in a preliminary measurement by STAR in peripheral isobar collisions (Fig.~\ref{fig:kaiyang}). The $\gamma$-induced $e^{+}e^{-}$ pairs which occur at low dielectron $p_{\mathrm{T}}$ are significantly (anti-)correlated with the event plane, in agreement with an EPA model prediction \cite{EPA-VMD}. This may allow for estimation of the reaction plane in heavy-ion collisions, as suggested in the previous reference. This result is being finalized for publication.\par 


\section{Conclusion and outlook}\label{sec:conclusion}

STAR has made significant contributions to our understanding of QCD in extreme conditions. In the vacuum, state-of-the-art models' description of the charge flow of hadronization has been tested. Observations of a non-linear increase in $J/\psi$ yields with multiplicity is suggestive of the influence of secondary partonic interactions or string percolation on $J/\psi$ production. In the medium, we have observed indications of path-length dependence of energy loss, although more work remains before making quantitative statements. Although jets deposit their energy in the medium, signals of medium-induced hadrochemistry effects and radius-dependence of yield modification are not observed in jets, within large uncertainties, placing an upper bound for the magnitude of effect that medium response can have in this kinematic region. The $J/\psi$ exhibits suppression of yield compared to pp collisions, even at center-of-mass energies on the order of $10 \mathrm{\ GeV}$, possibly due to a minimal contribution from regeneration in this regime. Additionally, sequential suppression of charmonium states has been observed for the first time at RHIC, which also serves as an indirect indication of the temperature of the thermalized medium. STAR has also reliably extracted the QGP temperature using thermal dielectrons, which is observed to be well above the pseudocritical temperature in the partonic phase, and near it in the later stages of the evolution. Finally, it has been observed that initial-state photon polarization and spin interference have an influence on the measured final-state collision products, giving access to initial state conditions and early-time dynamics in (ultra-)peripheral collisions.\par 
RHIC will finish data-taking at the end of 2025, with the last two Au+Au runs in 2023 and 2025 expected to deliver a three-fold increase in statistics for hard-probes measurements relative to the commonly-used 2014 Au+Au dataset. While these collisions are delivered, STAR will benefit from many upgrades from the past few years, including a new inner TPC, event plane detector, and forward upgrade, improved data acquisition rate, and more. With the resulting improved tracking precision, forward jet and heavy-flavor capabilities, and potential for unbiased centrality and event-plane determination, STAR will have a wealth of excellent data to analyze for the coming decade as the community prepares for the EIC.\par

\bibliography{STAR.bib} 
%

\end{document}